\documentclass[%
 reprint,
superscriptaddress,
 amsmath,amssymb,
 aps,
]{revtex4-2}

\usepackage{graphicx}% Include figure files
\usepackage{dcolumn}% Align table columns on decimal point
\usepackage{bm}% bold math
\usepackage{hyperref}% add hypertext capabilities
\begin{document}
%\preprint{APS/123-QED}

\title{Social influence on complex networks \\ as a perturbation to individual behavior}% Force line breaks with \\

\author{Dhruv Mittal}
\email{d.mittal@uva.nl}
\affiliation{Computational Science Lab, Informatics Institute, University of Amsterdam}%

\author{Flávio L. Pinheiro}\email{fpinheiro@novaims.unl.pt}
\affiliation{Nova Information Management School (NOVA IMS), Universidade Nova de Lisboa}%
\author{Vítor V. Vasconcelos}
\email{v.v.vasconcelos@uva.nl}
\affiliation{Computational Science Lab, Informatics Institute, University of Amsterdam}
\affiliation{POLDER, Institute for Advanced Study, University of Amsterdam}

\date{\today}% It is always \today, today,

\begin{abstract}
Cooperation is fundamental to the functioning of biological and social systems in both human and animal populations, with the structure of interactions playing a crucial role. Previous studies have used networks to describe interactions and explore the evolution of cooperation, but with limited transposability to social settings due to biologically relevant assumptions. Exogenous processes--- that affect the individual and are not derived from social interactions--- even if unbiased, have a role in supporting cooperation over defection, and this role has been largely overlooked in the context of network-based interactions. Here, we show that selection can favor either cooperation or defection depending on the frequency of exogenous, even if neutral, processes in any population structure. Our framework allows for deriving analytically the conditions for favoring a specific behavior in any network structure strongly affected by non-social environments (frequent exogenous forcing, FEF), which contrasts with previous computationally prohibitive methods. Our results demonstrate that the requirements for favoring cooperation under FEF do not match those in the rare-mutation limit, establishing that underlying neutral processes can be considered a mechanism for cooperation. We reveal that, under FEF, populations are less cooperative, and network heterogeneity can provide an advantage only if targeting specific network properties, clarifying seemingly contradictory experimental results and evolutionary predictions. While focused on cooperation, our assumptions generalize to any decision-making process involving a choice between alternative options. Our framework is particularly applicable to non-homogeneous human populations, offering a new perspective on cooperation science in the context of cultural evolution, where neutral and biased processes within structured interactions are abundant.

% \begin{description}
% Secondary publications and information retrieval purposes.
% You may use the \texttt{description} environment to structure your abstract;
% use the optional argument of the \verb+\item+ command to give the category of each item. 
% \end{description}
\end{abstract}

%Use showkeys class option if keyword
\maketitle

%\tableofcontents

\section{\label{sec:intro}Introduction\protect\\ }

Cooperation plays a fundamental role in the functioning of biological and social systems. Understanding how it emerges and prevails remains a significant challenge in biological and, especially, social disciplines \cite{hardin1998extensions, nowak2006five, sigmund2010calculus, santos2018social, hilbe2018evolution}. In biological systems, interactions are structured, e.g., through space and movement patterns, affecting evolutionary outcomes \cite{nowak2010evolutionary}. In social systems, online and offline social networks influence individuals' ability to coordinate in unanimously desirable outcomes \cite{vasconcelos2021segregation} and even the daily operation of social institutions.\cite{napoli2019social,tornberg2022digital,van2021social}. In that context, past works have addressed the evolution of cooperation using complex networks \cite{lieberman2005evolutionary} to describe interactions among individuals, but a common focus on biologically relevant assumptions limits the transposability of the results to social settings; namely, the assumption that events unrelated to interactions, including random changes in strategy (mutations in biological systems) and migration, are rare\cite{nowak2010evolutionary,ohtsuki2006simple,tarnita2009evolutionary,allen2017evolutionary}. 

In reality, social behaviors are highly variable in part due to the role of exogenous processes \cite{traulsen2010human}, or, more pragmatically, due to the multitude of factors affecting individuals' decisions \cite{albarracin2024determinants, tverskoi2023disentangling}. Furthermore, behavioral experiments demonstrate that evolutionary rules do not apply to human decisions in the lab. Instead, individuals have underlying cooperation preferences that are only partially influenced by their neighbors. \cite{grujic2014comparative} These external processes have been overlooked since, as we show here, even when unbiased, they can alter the conditions that indicate which behaviors are more frequent in specific network structures. In evolutionary terms, the selection of behaviors is a complex process that emerges from the interplay among the social structure, the specific interactions among individuals, and external factors to determine who is the fittest. 

Here, we demonstrate---for all population structures---that in evolutionary terms, selection can favor either cooperation or defection depending on the frequency of neutral exogenous processes, and that the conditions previously derived for the direction favored by selection are not universally applicable. We provide a closed-form perturbation formula for the behavioral response as a function of the linearization of the response function of individuals and the specific adjacency matrix describing the connections among individuals. For behavioral experiments, we show, for all possible network structures, that the response is either i) independent of topology with social influence contributing about 7\% to the average levels of cooperation or ii) shows a weak dependence on degree (0.5\% per additional connection per individual), depending on whether the responses obtained in experiments are stable for different fraction or number of neighbors' behavior. 

A key difference between past laboratory experiments and evolutionary settings is the dependencies of individual behavior being limited to first or second neighbors, respectively. We propose a foundational framework for studying social behavior and derive analytically the conditions that favor a specific behavior for any pairwise interaction in any network structure of nodes strongly influenced by their non-social environment, characterized by frequent exogenous forcing (FEF). We focus on cooperative behavior and further explore these conditions in evolutionary terms for arbitrary intensities of selection --- something that was previously thought to be computationally prohibitive \cite{ibsen2015computational, pinheiro2012selection} --- and the frequency of the exogenous forcing. For human behavior, as observed in the lab, we test different contexts for the behavioral rules and their implications for response in all networks. We demonstrate that the requirements for favoring cooperation under FEF do not align with those in the rare mutation limit, indicating that underlying neutral processes, or their absence, can be considered an evolutionary mechanism for cooperation. Moreover, our results show that, under FEF, populations are less cooperative, and network heterogeneity can only provide an advantage if it targets specific network properties –- further enlightening seemingly contradictory evolutionary predictions and experimental results.\cite{santos2005scale, gracia2012heterogeneous}. While we focus on cooperation and a specific update process, our assumptions about interactions generalize to any decision-making process involving a choice between alternative options, including those involving individuals with heterogeneous responses. Our framework applies more directly to diverse human populations where social interactions are but a contribution to behavioral change, and we expect our results to bring about a new perspective on cooperation science, especially in the context of human behavioral dynamics, where underlying neutral and biased processes within structured interactions abound.

\section{\label{sec:methods}Methods}
Overall, our goal is to understand how the link between (frequent) exogenous processes and social interactions in any network structure affects the evolutionary outcomes of behavior. But first, let us describe the general setup.

Consider a population of $N$ individuals who can be in one of two states and interact with each other in a network given by an adjacency matrix $A=[a_{ij}]$. In each time step, a single individual is uniformly chosen to update their strategy. In general, the probability the selected individual $i$ changes state is given by an update rule, $p_i^{s_i,\mathbf{s}}$, which depends on the individual ID, $i$, the individual’s current state, $s_i=\{0,1\}$ (standing for D and C, respectively), and the configuration of the population, $\mathbf{s} = \{s_1,...,s_N\}$, i.e., everyone’s state (we make the dependence on $s_i$ redundant for convenience). We assume that $p$ weakly depends on the interactions with others through a small parameter, $\gamma$, writing that dependency explicitly as $p_i^{s_i,\mathbf{s}} \equiv p_i^{(s_i)}[\gamma \phi_i^\mathbf{s}]$, where $\phi_i^\mathbf{s}$ contains the dependence on interactions with (potentially some of the) others. The fraction of individuals in state C, $x^\mathbf{s}:=\frac{1}{N}\sum_{i=1}^N s_i$, is a stochastic process and will change over time. 

%Average for any network structure and update rule (linearizable on social interactions)
We compute its expected value, in the long run, using the stationary distribution of the process, $P^\mathbf{s}[\gamma]$, as $\langle x \rangle := \sum_\mathbf{s} P^\mathbf{s}[\gamma]x^\mathbf{s}$. Expanding $P^\mathbf{s} [\gamma]$ in $\gamma$, that average is given by 
\begin{equation}
	\langle x \rangle = \langle x \rangle^{(0)} + \gamma \langle\Delta x\rangle^{(0)} + O(\gamma^2)
 \label{eq:averagex}
\end{equation}
where 
\begin{equation}
	\langle x \rangle^{(0)} \equiv \frac{1}{N}\sum^N_{i=1}\frac{p_i^{(0)}[0]}{p_i^{(0)}[0]+p_i^{(1)}[0]}
\end{equation}
and
\begin{widetext}
\begin{equation}
	\langle\Delta x\rangle^{(0)} \equiv \frac{1}{N} \sum_{i=1}^N\frac{1}{p_i^{(0)}[0]+p_i^{(1)}[0]}\left(p_i^{(0)'}[0]\sum_{\mathbf{s}:s_i=0}P_0^\mathbf{s}\phi_i^\mathbf{s}-p_i^{(1)'}[0]\sum_{\mathbf{s}:s_i=1}P_0^\mathbf{s}\phi_i^\mathbf{s}\right),
    \label{eq:Deltax_general}
\end{equation}
\end{widetext}
with $P^\mathbf{s}_0 \equiv P^\mathbf{s}[0]$ and $p_i^{(s_i)'}[0]= \left.\frac{d}{d x}p_i^{(s_i)}[x]\right\rvert_{x = 0}$ (see Appendix A for a detailed expansion). 

\begin{figure}[!t]
    \centering
    \includegraphics[width=0.45\textwidth]{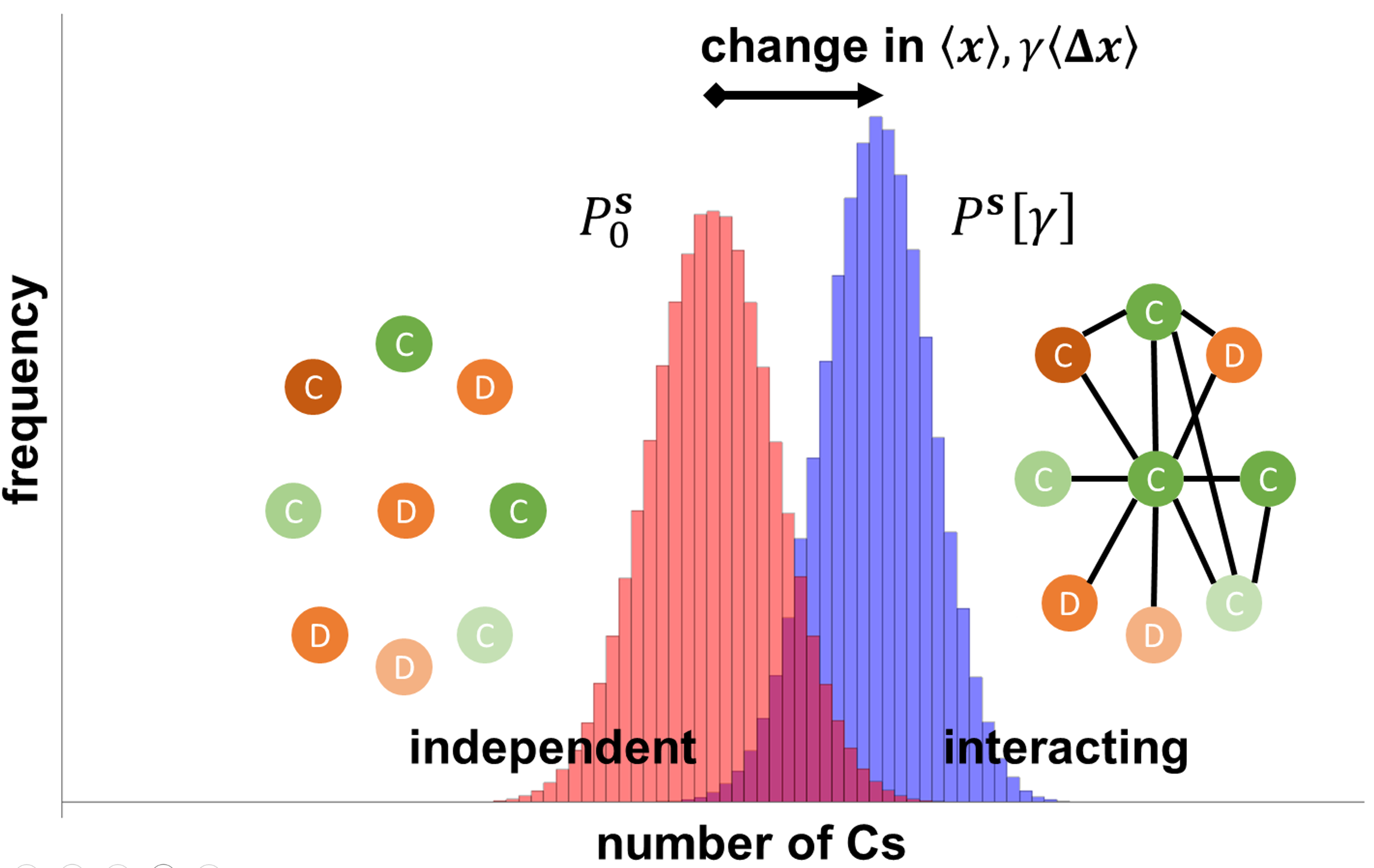}
    \caption{The frequent external forcing limit. In blue, we represent a perturbation of the red state in which all elements are independent of one another. As the interactions come into play via the network, the distribution shifts into one of two possible directions: interactions either favor the prevalence of state C or of state D. Individuals can be heterogeneous and the distribution does not need to be unimodal.}
    \label{fig:illustration}
\end{figure}

Figure \ref{fig:illustration} illustrates the dynamics captured by Eq.(\ref{eq:averagex}). The first term contains the expected fraction of individuals of type C when all individuals act independently of one another (red distribution, $P_0^\mathbf{s}$). In general, that is simply, $P_0^\mathbf{s}=\prod_{j=1}^N\alpha_j^{s_j}$, with $\alpha_j^{s_j}=p_j^{1-s_j}[0]⁄(p_j^{s_j} [0]+p_j^{1-s_j}[0])$. For the neutral case we discussed above, that would mean $p_j^{s_j}=p_j^{1-s_j}=\alpha_j^{s_j}=(P_0^\mathbf{s})^{1/N}=1/2$ for all $\mathbf{s}$. As we increase $\gamma$, the distribution shifts as a result of the networked interactions. Such a shift induces a change in the average fraction of one of the types, C in the illustrated case, given by $\gamma \langle\Delta x\rangle^{(0)}$, which corresponds to the second term of Eq.(\ref{eq:averagex}). The term $\langle\Delta x\rangle^{(0)}$ contains the effect that the network of interactions has on the expected value of the fraction of Cs and reflects the change of the distribution relative to its original shape. A positive $\langle\Delta x\rangle^{(0)}$ term indicates that selection, induced by the interactions among individuals, favors an increase in the abundance of state C and, for a negative term, it favors D. Notice how all terms depend on the model details (e.g., individual update rule), and the averages are taken in a distribution where $\gamma=0$, i.e., where there is no coupling between individuals. This gives a high degree of analytical tractability.

\section{Results}
\subsection{Analytical results for FEF and weak selection}
The island example we indicated above corresponds to the limit of high exploration (mutation) in evolutionary game theory, when individuals keep changing states, from cooperation (C, $s_i=1$) to defection (D, $s_i=0$) and back. For simplicity, when dealing with evolution in the remainder of this main text, we consider a pairwise imitation process given by the Fermi-update \cite{traulsen2007pairwise}. Later, we will focus on human behavior at faster scales where individual updates of states occur through "moody cooperation" that exhibits relatively small dependence on the number of cooperators around an individual, as empirical controlled experiments indicate \cite{grujic2014comparative}. 

When evolution occurs through pairwise imitation, a selected individual, $i$, changes strategy with probability $\mu$ due to exogenous factors and, with complementary probability $1-\mu$, selects one of its neighbors, $j$, to imitate based on the difference of expected payoffs. If the neighbor $j$ (defined such that $a_{ij}=1$) has a different strategy, the player changes strategy with a probability that depends on the difference of payoffs between the two individuals, $f_i^\mathbf{s}-f_j^\mathbf{s}$, given by $F[\beta(f_i^\mathbf{s}-f_j^\mathbf{s})]=(1+e^{\beta(f_i^\mathbf{s}-f_j^\mathbf{s})})^{-1}$. Thus,
\begin{equation}
	p_i^{s_i,\mathbf{s}}= \mu + (1-\mu)\sum_{j=1}^N\frac{a_{ij}}{k_i}\delta_{s_i(1-s_j)}F[\beta(f_i^\mathbf{s}-f_j^\mathbf{s})]
\end{equation}
is the probability that a selected individual changes strategy. While often used in the context of evolution, this update rule has strong relationships to human behavior. 
For almost two decades, strong linearities have been empirically demonstrated in the sociological context, where multiple reinforcement of a behavior is required by neighbors before the strategy change of the individual. \cite{centola2007complex, guilbeault2018complex} This process is coined as complex contagion and akin to the update described above, especially in the case of strong selection (high $\beta$), which makes the dependence on the interactions with neighbors captured by $f_i^\mathbf{s}-f_j^\mathbf{s}$ dominant.

\begin{figure*}[htp!]
    \centering
    \includegraphics[width=1\textwidth]{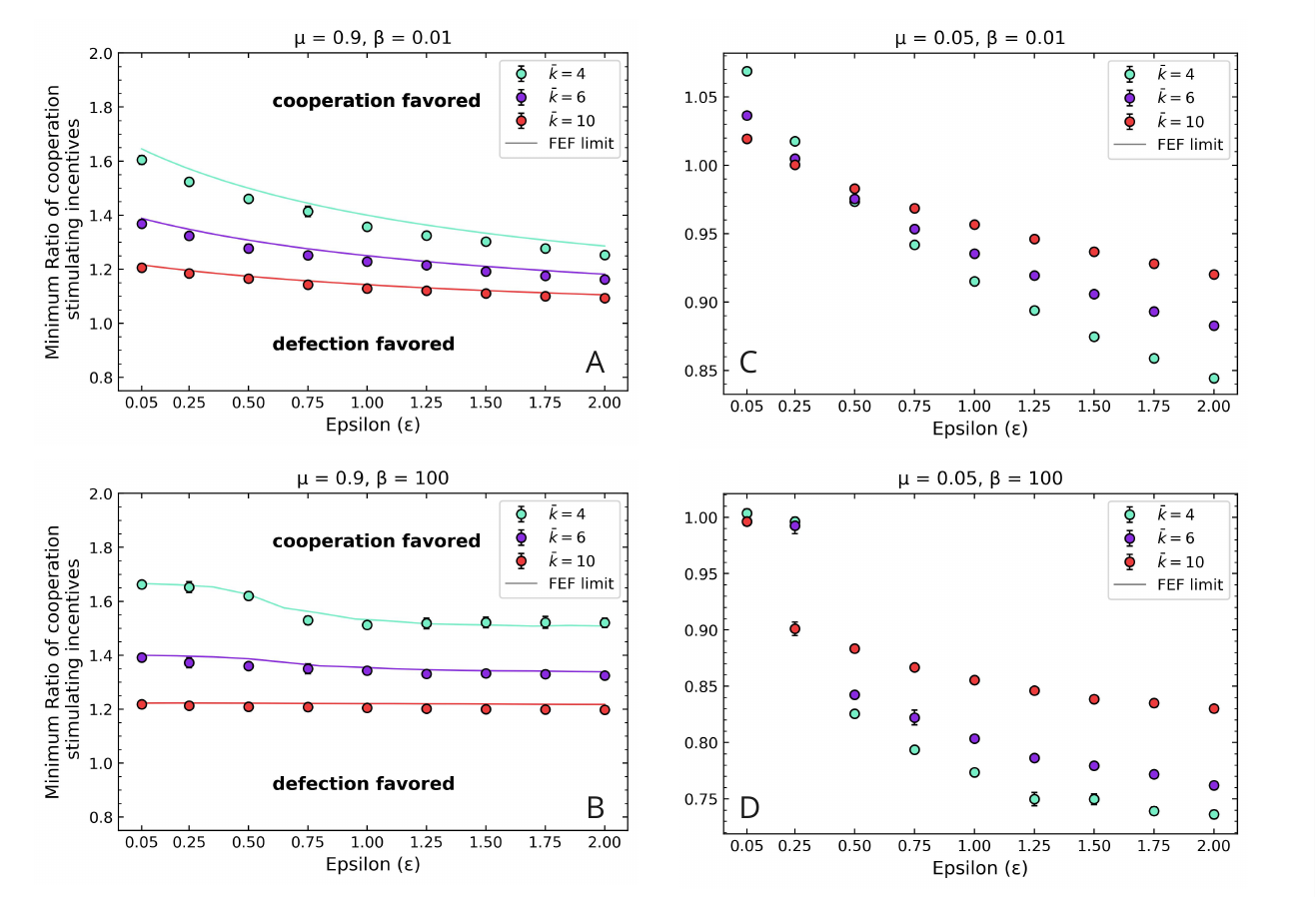}
    \caption{\textbf{Condition for cooperation to be modal in networks with different values of heterogeneity}, $\epsilon \equiv \bar{k_{nn}}/\bar{k} - 1$. The value of $\frac{R-P}{T-S}$ corresponding to an equilibrium of 50-50 cooperation and defection(let's call it $q$) calculated from simulations is plotted for such networks under varying values of selection intensity ($\beta$) and mutation rate ($\mu$). We also plot numerically calculated values of $q$ in the FEF limit using Eq.(\ref{eq:conditionFermi})) and using Eq.(\ref{eq:RPTS_condition}))( in the case of low ($\beta$)). In panels A and B, q is plotted for networks of different $\bar{k}$ and $\epsilon$ for low and high ($\beta$), respectively, in the FEF limit ($\mu=0.9$). In panels C and D, we plot q in the limit of rare mutation. Parameters: $P=0$, $R=1$, and, $S=1-T$. See methods for a description of how the plots were generated.}
    \label{fig:results1}
\end{figure*}
A standard framework to describe those interactions in the study of the evolution of cooperation is one in which individuals interact with their neighbors via a 2-player game. \cite{szabo2007evolutionary} There, if a cooperator interacts with a cooperator, they both get a reward, $R$.  If two defectors interact, they get a punishment payoff, $P$. If a cooperator interacts with a defector, the cooperator gets the sucker's payoff, $S$, and the defector gets $T$, temptation. 
Thus, the total payoff across all interactions of $i$ is just
\begin{widetext}
\begin{equation}
	f_i^\mathbf{s}=s_i \sum_{k=1}^N a_{ik}(Rs_k+S(1-s_k))+(1-s_i)\sum_{k=1}^Na_{ik}(Ts_k+P(1-s_k)).
\end{equation}
\end{widetext}
We find the condition for the number of cooperators to increase as interactions become more relevant in the players’ decisions by imposing that the second term in Eq.(\ref{eq:averagex}) above is positive, $\langle\Delta x\rangle^{(0)}>0$. In the limit of weak selection, we get---for any two-player game in any network---that
\begin{equation}
	\frac{\kappa -1}{\kappa + 1}R+S-T-\frac{\kappa -1}{\kappa + 1}P > 0
 \label{eq:conditionFermi}
\end{equation}
where $\kappa = \frac{\bar{k}+\bar{k_{nn}}}{2}$ and $\bar{k} \equiv \frac{1}{N} \sum_{i=1}^N k_i$ is the average degree of the network and $\bar{k}_{nn} = \frac{1}{N} \sum_{i=1}^N \frac{1}{k_i}\sum_{j=1}^N a_{ij}k_j$ is the mean average degree of the nearest neighbors. Notice that for homogeneous networks $\bar{k}_{nn} = \bar{k} = \kappa$. Otherwise, heterogeneous networks make $\kappa \neq \bar{k}$, where $\epsilon \equiv \frac{\bar{k}_{nn}}{\bar{k}} -1$ represents a measure of network heterogeneity. Another way of interpreting Eq.(\ref{eq:conditionFermi}) is to look at the impact of network heterogeneity on requiring a higher reward to mutual cooperation, $R$,  compared to mutual defection, $P$, relative to the temptation payoff, $T$, relative to the Sucker’s payoffs, $S$. When a higher ratio of $(R-P)/(T-S)$ is required for cooperation to be favored, it means cooperation is harder (cooperation requires higher $R$ or $S$ or lower $T$ or $P$). Eq.(\ref{eq:conditionFermi}) can be rewritten, for $T>S$, as 
\begin{equation}
	\frac{R-P}{T-S} > \frac{\kappa+1}{\kappa-1} = \frac{\bar{k}(1+\epsilon/2)+1}{\bar{k}(1+\epsilon/2)-1},
    \label{eq:RPTS_condition}
\end{equation}
which is a decreasing function of heterogeneity, $\epsilon$. Thus, for the pairwise comparison update rule, heterogeneity facilitates cooperation when exogenous events are common. While Eq.(\ref{eq:Deltax_general}) is valid for any network, agents with heterogeneous update rules, including non-linear ones (non-weak selection), Eqs.(\ref{eq:conditionFermi},\ref{eq:RPTS_condition}) hold for any network structure but only under homogeneous, linear updates (weak selection), which can also be derived using the McAvoy-Wakeley method. \cite{mcavoy2022evaluating} Importantly, the condition in Eq.(\ref{eq:conditionFermi}) identifies a simple, unique network property that governs cooperation in the limit of strong exogenous forcing, $\kappa$, for all networks and pairwise interactions. 

With the framework proposed, besides computing the analytical conditions for cooperation under weak selection, we can efficiently solve this problem numerically for arbitrary selection intensities, whereas, before, no such algorithm was known \cite{ibsen2015computational}.

\subsection{Dependence on frequency of exogenous events and selection intensity}

 We now compare the payoff conditions measured using the ratio $(R-P)/(T-S)$ required for cooperation to be more prevalent for varying intensities of selection and mutation rates across different network structures. 

Fig. \ref{fig:results1}A, B show that payoff conditions above which cooperation is favored, in the FEF limit for weak and strong selection, respectively. The values calculated from Monte Carlo simulations match Eq.\ref{eq:RPTS_condition} in the case of weak selection and the numerical calculations using Eq.\ref{eq:Deltax_general}, for strong selection. In both cases, cooperation is only favored in a Harmony Game. In contrast, for a low mutation rate ($\mu\rightarrow 0$), cooperation can be favored in a Prisoners' Dilemma for more heterogeneous networks (Fig.\ref{fig:results1} C, D). We see a clear departure from the condition obtained in the previous section.

Network heterogeneity improves the payoff conditions required to favor cooperation across different limits of mutation rate and selection intensity. However, for strong selection in the FEF limit, this effect is less pronounced, especially for networks with higher average degree (Fig. \ref{fig:results1}B). Further in the FEF limit, a higher average degree of the network, $\bar{k}$, improves the payoff conditions required for greater cooperation regardless of the network heterogeneity.  This is a clear departure from the limit where mutation is rare. For a low mutation rate, a lower average degree is conducive to cooperation except for relatively homogeneous networks \cite{ohtsuki2006simple}.

\begin{figure*}[htp!]
    \centering
    \includegraphics[width=\textwidth]{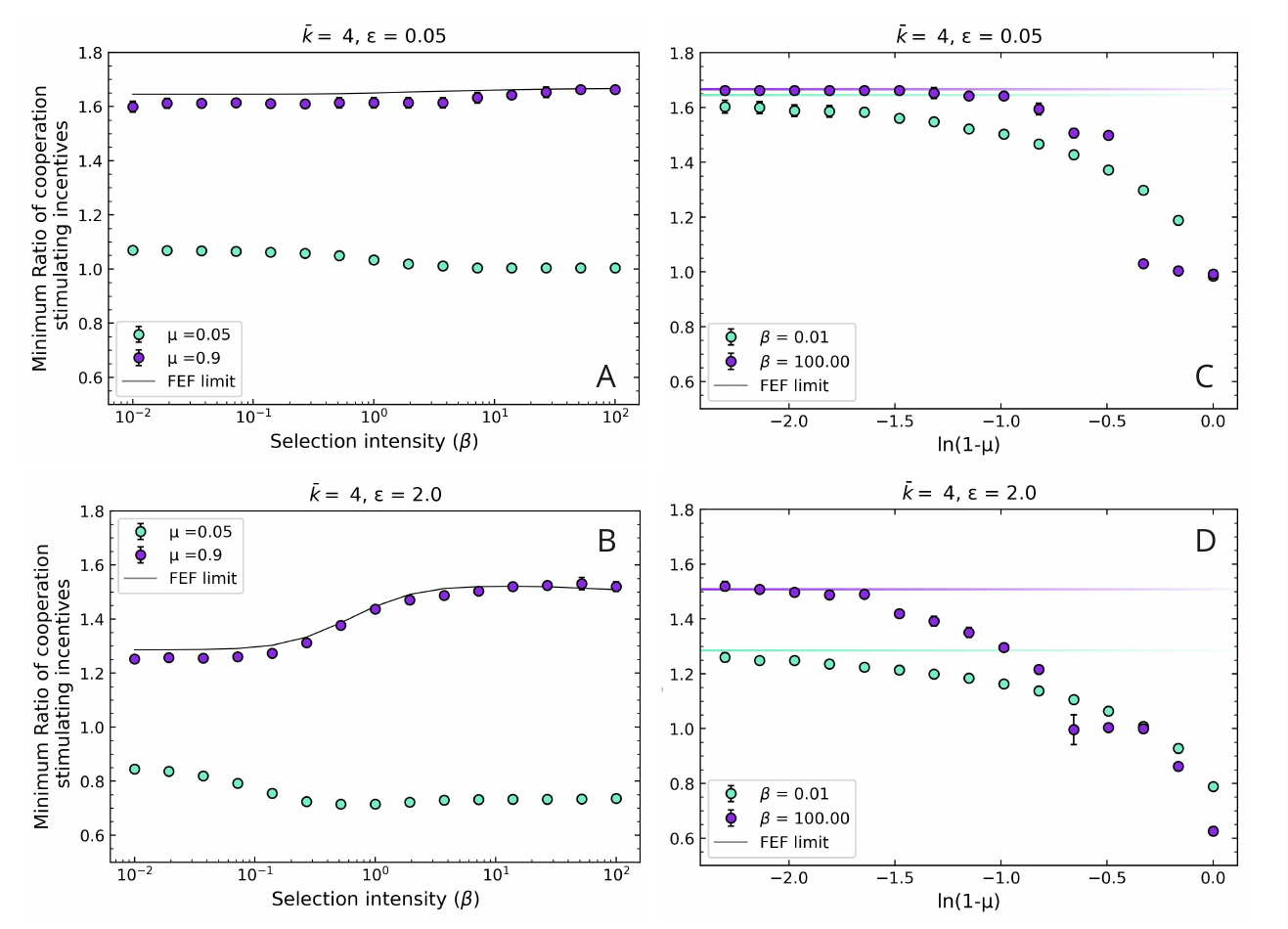}
    \caption{\textbf{Condition for cooperation to be modal in networks under varying selection pressure and mutation rate}, $\epsilon \equiv \bar{k_{nn}}/\bar{k} - 1$. The value of $\frac{R-P}{T-S}$ corresponding to an equilibrium of 50-50 cooperation and defection(let's call it $q$) calculated from simulations is plotted for such networks under varying values of selection intensity ($\beta$) and mutation rate ($\mu$). We also plot numerically calculated values of $q$ in the FEF limit using Eq.(\ref{eq:conditionFermi})). In panels A and B, we plot q under varying $\beta$ for homogeneous and heterogeneous networks in opposite limits of the mutation rate. In panels C and D,  the results are shown similarly with $\mu$ being varied. Parameters: $P=0$, $R=1$, and, $S=1-T$. See methods for a description of how the plots were generated.}
    \label{fig:results2}
\end{figure*}

 Next, we calculate the conditions that favor cooperation for varying selection intensity. In both limits of $\mu$, in the weak selection approximation, the effect of interactions is weak and acts on a symmetric distribution with a mean of $1/2$. However, the clear difference (see Figure \ref{fig:results2}A, B)in outcomes highlights the importance of non-payoff driven processes in changing a population's evolutionary outcome, even if those processes are neutral in favoring one strategy vs. the other. \cite{mcavoy2022evaluating} Our analysis further shows that this dependency holds even for higher levels of selection intensity. Selection favors cooperation in the absence of exploration, but it favors defection when exploration is frequent, both for strong- and weak-selection regimes. This divergence is more pronounced for heterogeneous networks. Furthermore, we also observe that for heterogeneous networks and a high rate of mutation, cooperation is not favored in the prisoner's dilemma at any level of selection pressure.

Figures \ref{fig:results2}C, D show how the condition varies as a function of the mutation rate. We can see how the conditions for selection to favor cooperation are more consistent for higher rates of exogenous forcing than when exogenous forcing is rare. This brings hope to understanding the impact of network structure on behavior in human populations, since those are expected to be dominated by exogenous events. In this limit, conditions for cooperation are not as sensitive to the exact exogenous rate or the level of errors captured by the intensity of selection.

\subsection{Behaviors in human experiments of cooperative behavior}
The decades of analytical sociological study of social innovations that "face resistance---because they are unfamiliar or difficult, or because they challenge existing social norms" \cite{centola2021complex}---emphasize the importance of reinforcement, non-linear response to social environments, and the role of heterogeneous networks that we just discussed. However, a different behavior has been identified when it comes to behaviors that do not face resistance.

In human experiments of cooperative behavior, behavioral responses are at shorter timescales and do not face the resistance that is implicit in threshold-like update rules. While playing against others in experiments with repeated interactions, humans exhibit conditional cooperation strategies, or 'moody cooperation.' \cite{grujic2014comparative} Under those circumstances, the behavior is also characterized by a weak dependence on the behavior of the neighborhood, with a significant part of the probability of cooperating being relatively stable across different neighborhoods. Using our framework, we can predict the levels of cooperation for any structure (see Appendix E).

In the experiments, the authors report the behavioral rules distilled for a specific game played in those experiments \cite{grujic2010social}. There, they report the response of individuals with different strategies to the number of cooperators in their neighborhood. Thus, that dependence can be seen as a function of the number of individuals or as a function of the fraction. When we consider that individuals respond to the number of cooperators around then, we can use our framework to calculate the expected level of cooperation using our formula as $\langle x \rangle = 0.284+\gamma (0.0216+0.00589 \langle k\rangle)+O(\gamma^2)$, where $\langle k\rangle$ is the average degree of the network and $\gamma=1$ matches the individual behavior in the experiment. This expression shows that denser networks have a higher level of cooperation for the same game and set the influence of the neighbors. Further, it shows that the effect of the peers is positive. However, this is not the only model that is possible at the individual level.  When considering the fraction, the effect of network density on overall cooperation disappears, and we predict a fixed level of cooperation at $\langle x \rangle = 0.284 + 0.0687 \gamma + O(\gamma^2)$. In that case, the effect of peer influence is still positive but independent of the network structure.
These results highlight how different update processes change the impact of the social network on favoring different behaviors. In particular, under the frequent exogenous processes limit, update responses that have no dependency on second-degree neighbors do not exhibit dependency on the heterogeneity of the network. In the Fermi update, that dependency comes from the payoff comparison, in which the payoff of the first-degree neighbor is directly dependent on the strategy of their neighbors. 

%----------------------------------------------------------

\section{Conclusions and discussion} 
Understanding the effect networks have on human behavior is vital for modern society. Though a complex problem, it has now been solved for any graph structure, both for rare and common underlying exogenous processes, like mutation and migration. Our work shows that selection acts in more complex ways than anticipated, and its effect on cooperation is strikingly different when it is the main decision driver versus when it is the sole component of individuals’ evolution. This result asks for a profound reassessment of the transposability of the results derived for cooperation in biological systems with close to null mutation rates to those where mutation and migration are active elements, in particular in what concerns the effect of the structure of interactions. The average degree of the network, $\bar{k}$, had been identified as determining the conditions for higher cooperation in the case of networks of homogeneous degree networks, where $\kappa$ and $\bar{k}$ match.\cite{ohtsuki2006simple} However, it fails to predict behavior in heterogeneous networks.\cite{allen2024coalescent,santos2008social}. We show that in the high mutation limit specific network property can determine if the structure favors cooperation.

In this study, we varied the selection strength and mutation rate to investigate how the steepness, as well as linear transformations of the response function, interact with the underlying complex network structures that shape the evolutionary dynamics. In the case of FEF, the social influence can be treated as a perturbation, allowing us to study the system without prohibitive costs of computation. In social systems, we can interpret selection pressure as rationality and mutation rate as the rate of exploration. 

Other studies, focused on evolutionary processes, have also found that high mutation rates inhibit the emergence of cooperation in conditions under which it would emerge \cite{ichinose2018mutation, takesue2019roles, allen2012mutation}. However, the rationality and network heterogeneity can affect cooperation positively or negatively to different extents. With our approach, which accounts for payoff, network structure, rationality, and exploration, we show that this is, indeed, not universal. In homogeneous networks, varying selection pressure has little effect on the payoff conditions that favor cooperation. In contrast, in heterogeneous networks, the interaction between mutation and selection pressure produces opposing outcomes depending on the exploration rate. Under high mutation rates, stronger selection makes cooperation harder to sustain, but, when mutations are rare, higher selection pressure creates more favorable conditions for cooperation, likely due to the disproportionate influence of highly connected hubs. Overall, while network heterogeneity promotes cooperation, this advantage is significantly weakened with high rationality.

While we have looked at static structures in this study, it remains unclear if the conditions for cooperation derived here would also hold for dynamic structures with similar properties. This question could be particularly relevant if heterogeneous structures lead to local coordination, creating path dependence that favors either choice. 

The framework we propose and results in Eq.(\ref{eq:averagex}) is general in scope, allowing one to compute the conditions for cooperation with many different assumptions. Those include, e.g., situations in which the population is heterogeneous and individuals have different update rules; extending 2-player games to $n$-player interactions; and considering other reproduction processes.

Further, our analysis of the effect of mutation and rationality has focused on games within the spectrum from the Harmony game to the Prisoner’s Dilemma. Expanding the scope to encompass the full range of two-player games and beyond would provide a more complete understanding of cooperative dynamics across different levels of exploration and selection intensity.

In the appendix, we provide some forms of this equation that are particular to the cumulative assumptions we used in deriving the condition in Eq. (\ref{eq:conditionFermi}), but we, by no means, aimed at exploring here all the potential it provides in solving many open questions. From the setup, it is clear that the FEF limit can be used, for instance, to thoroughly study update rules that are not based on contact but in self-evaluation, such as aspiration dynamics\cite{simon1955behavioral,roca2011emergence,du2014aspiration} and best responders or gradient climbers\cite{matsui1992best, roca2009promotion, hu2013cooperative, ogren2004cooperative} as those, in the limit of weak selection, are comprised of independent agents. The way that dependency is created is a function of the heterogeneity of the individuals and can potentially be expressed in closed form using this formalism. 
Furthermore, we believe that, if the study of cooperation in human societies has learned so much from the techniques and models stemming from biological systems, and their consequent typical assumptions, it is only natural that biological systems can also obtain new insights inspired by the dynamics of social systems, and their assumptions, such as the FEF limit considered here.

\section*{Methods}
\noindent \textbf{Monte Carlo Simulations} have been used to validate and complement the analytical and numerical predictions obtained from equations. For each value of $\bar{k}$ and $\epsilon$, we generate 100 networks. For each set of $\mu$ and $\beta$ values, we run a total of 100 independent simulations, each starting with a random composition of strategies. At each step in the Monte Carlo simulations, a random individual($j$) imitates a randomly chosen neighbor with probability given by the Fermi function $(1-\mu)(1+e^{\beta(f_i^\mathbf{s}-f_j^\mathbf{s})})^{-1}$. With probability $\mu$, the same individual changes strategy randomly. For each simulation, we measure the observed balance of strategies, i.e., the equilibrium fraction of cooperators ($x^*$), by averaging over 4000 generations after a transient of 2000 generations. 

\noindent \textbf{Calculating condition favoring cooperation}

\noindent We perform this analysis for the family of games defined by $P=0$, $R=1$, and $S=1-T$. To determine the critical value $T_\text{crit}$ at which the mean value of $x^*$ equals $0.5$, we first compute the mean and standard deviation of $x^*$ across all realizations for a linearly spaced list of $T$ values. We then iteratively narrow the interval of $T$ by moving inward from both ends of the $T$ range, discarding values where the 95\% confidence interval (CI) of $x^*$ (estimated as $\bar{x}^*_T \pm 1.96\, \sigma_T$) lies entirely above or below $0.5$. The remaining interval, $[T_{\text{lower}}, T_{\text{upper}}]$, provides a CI for $T_\text{crit}$, and these values of $T_{\text{lower}}$ and $T_{\text{upper}}$ are used as updated range limits for the next iteration to progressively refine the estimate. The midpoint $(T_{\text{lower}} + T_{\text{upper}})/2$ is taken as the best estimate of $T_\text{crit}$ for that iteration.

If this procedure cannot narrow the CI for $T_\text{crit}$ to the desired level, we fit a linear regression of $x^*_T$ versus $T$ within the remaining range to estimate $T_\text{crit}$ for each network realization. The mean of these estimates across all realizations is taken as the best estimate of $T_\text{crit}$, while the 95\% CI of these estimates defines the new range limits of $T$ for the next iteration. This hybrid approach ensures robust estimation of $T_\text{crit}$ even when $x^*_T$ is nonlinear in $T$. The initial discarding step reduces the range to where $x^*_T$ is approximately linear, enabling the linear regression step to pool data across $T$ for a more precise estimate of $T_\text{crit}$, if needed. Finally, we calculate the ratio $(R-P)/(T-S)$, which simplifies to $1/(2T-1)$, for $T_\text{crit}$ and its CI, as shown in Figures~\ref{fig:results1}, \ref{fig:results2}.

\noindent \textbf{Networks} with different levels of $\epsilon$ have been generated through an algorithm of biased rewiring. Starting from a homogeneous random graph, with average degree $\bar{k}$ and $\epsilon \approx 0$, links are iteratively rewired, but only the rewires that increase the value of $\epsilon$, the algorithm stops when the desired level of $\epsilon$ is reached. 

\subsection*{Data Availability}
%The data that support the findings of this study are available from the corresponding author upon reasonable request.	
All simulation data is generated with the code available and described in the repository (\url{https://doi.org/10.5281/zenodo.17182768}). 

\begin{acknowledgments}
Vítor V. Vasconcelos would like to thank Lei Zhou for the discussions and developments made regarding aspiration dynamics, which were the inspiration for the development of this project.
Flávio L. Pinheiro is thankful for the funding from FCT (Fundação para a Ciência e a Tecnologia), under the project UIDB/04152 - Centro de Investigação em Gestão de Informação (MagIC)/NOVA IMS.

\end{acknowledgments}

% \bibliography{apssamp}% Produces the bibliography via BibTeX.

% \clearpage

\appendix
\begin{widetext}
%\section*{General Framework}
\section{General Framework}

Consider a population with $N$ individuals. Each individual, $i \in \{1, \ldots, N\}$, can be in one of two states, $B$ or $A$, corresponding to $s_i = \{0,1\}$, respectively. At any point in time, the evolution of the system is determined by the configuration of the population $\mathbf{s} = \{s_1, s_2, \ldots, s_N\}$. Thus, the set of states of all individuals indexed by time is a Markovian stochastic process. In each time step, an individual can change state according to an update rule that depends on its current state, the population configuration, and eventually its own individual properties, $p_{s_i, i}^{\mathbf{s}}$. We assume that the update function has the form
\[
p_{s_i, i}^{\mathbf{s}} \equiv p_i(s_i)[\gamma \Phi_i^{\mathbf{s}}],
\]
where $\gamma$ is a small parameter of this update rule and $\Phi_i^{\mathbf{s}}$ contains the dependence on the configuration of the system, $\mathbf{s}$.

Let $x_{\mathbf{s}}$ be the fraction of individuals in state $A$ ($s_i=1$), given by
\[
x_{\mathbf{s}} = \frac{1}{N} \sum_{i=1}^N s_i.
\]
Let $P_{\mathbf{s}}[\gamma]$ represent the stationary occupation of configuration $\mathbf{s}$, which depends on $\gamma$. The average (stationary) fraction of $A$'s is given by

%\begin{widetext}
\begin{equation}
\langle x \rangle = \sum_{\mathbf{s}} P^{\mathbf{s}}[\gamma] x^{\mathbf{s}}
= \sum_{\mathbf{s}} P_0^{\mathbf{s}} x_{\mathbf{s}}
+ \gamma \sum_{\mathbf{s}} P_{(1)}^{\mathbf{s}} x^{\mathbf{s}}
+ O(\gamma^2),
\end{equation}
%\end{widetext}

where $P_0^{\mathbf{s}} = P^{\mathbf{s}}[0]$ and $P_{(1)}^{\mathbf{s}} = \left.\frac{d P^{\mathbf{s}}}{d\gamma}\right|_{\gamma=0}$.

In what follows, we prove that

%\begin{widetext}
\begin{flalign*}
\langle x \rangle &= \frac{1}{N} \sum_{i=1}^N \frac{p_i^{(0)}[0]}{p_i^{(0)}[0] + p_i^{(1)}[0]} \\
&\quad + \gamma \frac{1}{N} \sum_{i=1}^N \frac{1}{p_i^{(0)}[0] + p_i^{(1)}[0]}
\left(
p_i^{(0)\prime}[0] \sum_{\mathbf{s}: s_i = 0} P_0^{\mathbf{s}} \Phi_i^{\mathbf{s}}
- p_i^{(1)\prime}[0] \sum_{\mathbf{s}: s_i = 1} P_0^{\mathbf{s}} \Phi_i^{\mathbf{s}}
\right)
+ O(\gamma^2). &&
\end{flalign*}

%\end{widetext}

Notice that the sums over the states are fairly simple to compute, given that they are done over a distribution in which the states of individuals are independent of each other. In the next section, we illustrate how this formulation can be used in traditional problems of Evolutionary Game Theory. In the following section, we look at transition probabilities that derive directly from experimental data on human behavior.

\subsection{Computation of the first term}

Due to the way we define $\gamma$, for $\gamma = 0$, all updates of a single individual are independent of the states of the remaining individuals. Then, it follows that $P_0^{\mathbf{s}}$ can be decomposed into the product of $N$ independent probability distributions corresponding to each individual changing independently between the two states. In this case, for each individual, the probability distributions are those of a Markov chain with only two states and constant transition probabilities. Thus, for the whole population,
\[
P^0_{\mathbf{s}} = \prod_{i=1}^N \alpha_i^{(s_i)},
\]
with
\[
\alpha_i(s_i) = \frac{p_i^{(1-s_i)}[0]}{p_i^{(s_i)}[0] + p_i^{(1-s_i)}[0]}.
\]
This immediately gives the first term in Eq.(1):

%\begin{widetext}
\begin{align}
\sum_{\mathbf{s}} P_0^{\mathbf{s}} x_{\mathbf{s}} 
&= \sum_{\mathbf{s}} \left( \prod_{j=1}^N \alpha_j^{(s_j)} \right) \left( \frac{1}{N} \sum_{i=1}^N s_i \right)
= \frac{1}{N} \sum_{i=1}^N \sum_{\mathbf{s}} \left( \prod_{j=1,\, j \neq i}^N \alpha_j^{(s_j)} \right) \alpha_i^{(s_i)} s_i \notag\\
&= \frac{1}{N} \sum_{i=1}^N \alpha_i^{(1)} \sum_{\mathbf{s}: s_i = 1} \prod_{j \neq i} \alpha_j^{(s_j)} 
= \frac{1}{N} \sum_{i=1}^N \alpha_i^{(1)}.\notag
\end{align}
%\end{widetext}

\subsection{Computation of the second term}

The computation of the second term in Eq.(1) is more complicated. Let us define
\[
\Omega_i \equiv \sum_{\mathbf{s}: s_i = 1} P_{(1)}^{\mathbf{s}}.
\]
Then, we can write the second term for the average fraction of cooperators, Eq.(1), as:
\[
\gamma \sum_{\mathbf{s}} P_{(1)}^{\mathbf{s}} x_{\mathbf{s}}
= \gamma \sum_{\mathbf{s}} P_{(1)}^{\mathbf{s}} \left( \frac{1}{N} \sum_{i=1}^N s_i \right)
= \gamma \frac{1}{N} \sum_{i=1}^N \sum_{\mathbf{s}} P_{(1)}^{\mathbf{s}} s_i
= \gamma \frac{1}{N} \sum_{i=1}^N \Omega_i. \tag{A2}
\]

In what follows, we show that
\[
\Omega_i = \frac{1}{p_i^{(0)}[0] + p_i^{(1)}[0]}
\left(
p_i^{(0)\prime}[0] \sum_{\mathbf{s}: s_i = 0} P_0^{\mathbf{s}} \Phi_i^{\mathbf{s}}
- p_i^{(1)\prime}[0] \sum_{\mathbf{s}: s_i = 1} P_0^{\mathbf{s}} \Phi_i^{\mathbf{s}}
\right),
\]
which proves the expression for the second term.

\subsection{Proof}

To compute \( \Omega_i \equiv \sum_{\mathbf{s}} P_{(1)}^{\mathbf{s}} s_i \), we make use of the master equation for the whole Markov chain corresponding to the states of each node. Let \( T^{\mathbf{s}' \to \mathbf{s}}[\gamma] \) represent the one-step transition probability that the system goes from configuration \( \mathbf{s}' \) to configuration \( \mathbf{s} \). 

By definition of transition probability:
\[
\sum_{\mathbf{s}} T^{\mathbf{s}' \to \mathbf{s}}[\gamma] = 1,
\]
and only “one-step transitions” are non-zero, where \( \mathbf{s}' \) and \( \mathbf{s} \) differ by at most one entry (e.g., \( \mathbf{s}' = \{0,1,0,0,\ldots\} \), \( \mathbf{s} = \{0,0,0,0,\ldots\} \)).

In such a case,
\[
T^{\mathbf{s}' \to \mathbf{s}}[\gamma] = \frac{1}{N} p_{s_i,i}^{\mathbf{s}'} = \frac{1}{N} p_i^{(s_i)}[\gamma \Phi_i^{\mathbf{s}'}].
\]

Using the system’s steady state equation and expanding in \( \gamma \), we get for \( P_{(1)}^{\mathbf{s}} \):
\[
P^{\mathbf{s}}[\gamma] = \sum_{\mathbf{s}'} P^{\mathbf{s}'}[\gamma] T^{\mathbf{s}' \to \mathbf{s}}[\gamma]
\Rightarrow
P_{(1)}^{\mathbf{s}} = \sum_{\mathbf{s}'} P_{(1)}^{\mathbf{s}'} T^{\mathbf{s}' \to \mathbf{s}}[0]
+ \sum_{\mathbf{s}'} P_0^{\mathbf{s}'} T_{(1)}^{\mathbf{s}' \to \mathbf{s}}, \tag{A3}
\]
with
\[
T_{(1)}^{\mathbf{s}' \to \mathbf{s}} = \left. \frac{d}{d\gamma} T^{\mathbf{s}' \to \mathbf{s}}[\gamma] \right|_{\gamma = 0}.
\]

From conservation of probability:
\[
\sum_{\mathbf{s}} P^{\mathbf{s}}[\gamma] = 1 \Rightarrow 0 = \sum_{\mathbf{s}: s_i = 0} P_{(1)}^{\mathbf{s}} + \Omega_i
\Rightarrow -\Omega_i = \sum_{\mathbf{s}: s_i = 0} P_{(1)}^{\mathbf{s}}. \tag{A4}
\]

Also, the one-step transitions give:
\[
\sum_{\substack{\mathbf{s}: s_i = 1, \\ \mathbf{s}': s_i' = 1}} T^{\mathbf{s}' \to \mathbf{s}}[\gamma]
= 1 - \frac{1}{N} p_i^{(1)}[\gamma \Phi_i^{\mathbf{s}'}], \tag{A5a}
\]
\[
\sum_{\substack{\mathbf{s}: s_i = 1, \\ \mathbf{s}': s_i' = 0}} T^{\mathbf{s}' \to \mathbf{s}}[\gamma]
= \frac{1}{N} p_i^{(0)}[\gamma \Phi_i^{\mathbf{s}'}]. \tag{A5b}
\]

Expanding the sums in Eq.(5) for all \( \mathbf{s} \) with \( s_i = 1 \):

\begin{align*}
\Omega_i &= \sum_{\mathbf{s}': s_i' = 0} P^{\mathbf{s}'}_{(1)} \sum_{\mathbf{s}: s_i = 1} T^{\mathbf{s}' \to \mathbf{s}}_{[0]}
+ \sum_{\mathbf{s}': s_i' = 1} P^{\mathbf{s}'}_{(1)} \sum_{\mathbf{s}: s_i = 1} T^{\mathbf{s}' \to \mathbf{s}}_{[0]}  \\
&\quad + \sum_{\mathbf{s}': s_i' = 0} P^{\mathbf{s}'}_{0} \sum_{\mathbf{s}: s_i = 1} T^{\mathbf{s}' \to \mathbf{s}}_{(1)}
+ \sum_{\mathbf{s}': s_i' = 1} P^{\mathbf{s}'}_{0} \sum_{\mathbf{s}: s_i = 1} T^{\mathbf{s}' \to \mathbf{s}}_{(1)} \\
&= \sum_{\substack{\mathbf{s}: s_i = 1 \\ \mathbf{s}': s_i' = 0}} P^{\mathbf{s}'}_{(1)} T^{\mathbf{s}' \rightarrow \mathbf{s}}_{[0]}
+ \sum_{\substack{\mathbf{s}: s_i = 1 \\ \mathbf{s}': s_i' = 1}} P^{\mathbf{s}'}_{(1)} T^{\mathbf{s}' \rightarrow \mathbf{s}}_{[0]}
+ \sum_{\substack{\mathbf{s}: s_i = 1 \\ \mathbf{s}': s_i' = 0}} P^{\mathbf{s}'}_{0} T^{\mathbf{s}' \rightarrow \mathbf{s}}_{(1)}
+ \sum_{\substack{\mathbf{s}: s_i = 1 \\ \mathbf{s}': s_i' = 1}} P^{\mathbf{s}'}_{0} T^{\mathbf{s}' \rightarrow \mathbf{s}}_{(1)} \\
&= \sum_{\mathbf{s}': s_i' = 0} P^{\mathbf{s}'}_{(1)} \sum_{\mathbf{s}: s_i = 1} T^{\mathbf{s}' \rightarrow \mathbf{s}}_{[0]}
+ \sum_{\mathbf{s}': s_i' = 1} P^{\mathbf{s}'}_{(1)} \sum_{\mathbf{s}: s_i = 1} T^{\mathbf{s}' \rightarrow \mathbf{s}}_{[0]} \\
&\quad + \sum_{\mathbf{s}': s_i' = 0} P^{\mathbf{s}'}_{0} \sum_{\mathbf{s}: s_i = 1} T^{\mathbf{s}' \rightarrow \mathbf{s}}_{(1)}
+ \sum_{\mathbf{s}': s_i' = 1} P^{\mathbf{s}'}_{0} \sum_{\mathbf{s}: s_i = 1} T^{\mathbf{s}' \rightarrow \mathbf{s}}_{(1)} \\
&= -\Omega_i \cdot \frac{1}{N} p^{(0)}_i[0] + \left(1 - \frac{1}{N} p^{(1)}_i[0]\right)\Omega_i
+ \frac{1}{N} p^{(0)\prime}_i[0] \sum_{\mathbf{s}': s_i' = 0} P^{\mathbf{s}'}_{0} \Phi_i^{\mathbf{s}'} \\
&\quad - \frac{1}{N} p^{(1)\prime}_i[0] \sum_{\mathbf{s}': s_i' = 1} P^{\mathbf{s}'}_{0} \Phi_i^{\mathbf{s}'} \\
&= \frac{1}{p^{(0)}_i[0] + p^{(1)}_i[0]} \left(
p^{(0)\prime}_i[0] \sum_{\mathbf{s}: s_i = 0} P^{\mathbf{s}}_{0} \Phi_i^{\mathbf{s}}
- p^{(1)\prime}_i[0] \sum_{\mathbf{s}: s_i = 1} P^{\mathbf{s}}_{0} \Phi_i^{\mathbf{s}}\right). \tag{A6}\\
&\qquad \qquad \qquad \qquad \qquad \square
\end{align*}

\section{The weak dependence limit in evolutionary game theory – the Fermi update}

Let us assume that a population evolves according to a (single) pairwise update process with (homogeneous) mutation in a network with adjacency matrix $a_{ij}$ (no self-links). The probability that any selected individual updates their strategy is given by

\[
p^{\mathbf{s}}_{s_i,i} = \mu + (1 - \mu) \sum_{j=1}^N \frac{a_{ij}}{k_i} \delta_{s_i(1 - s_j)} F[\beta(f^{\mathbf{s}}_i - f^{\mathbf{s}}_j)]\tag{A7}
\]

Thus, if we set $\gamma = 1 - \mu$, and look in the limit $\mu \sim 1$, the system obeys the general framework described in the first section with

\[
\Phi_i^{\mathbf{s}} = -1 + \sum_{j=1}^N \frac{a_{ij}}{k_i} \delta_{s_i(1 - s_j)} F[\beta(f^{\mathbf{s}}_i - f^{\mathbf{s}}_j)]
\]

Then, we can write $p^{s_i}_i[\gamma \Phi_i^{\mathbf{s}}] = 1 + \gamma \Phi_i^{\mathbf{s}}$, making $p^{(0)}_i[0] = p^{(1)}_i[0] = 1$, $p^{(0)\prime}_i[0] = p^{(1)\prime}_i[0] = 1$. Under those assumptions, Eq.(9) becomes

\[
\langle x \rangle = \frac{1}{2} + (1 - \mu) \mathcal{B} + \mathcal{O}((1 - \mu)^2) 
\]

with

\[
\mathcal{B} = \frac{1}{2N} \sum_{i=1}^N \left( \sum_{\mathbf{s}: s_i = 0} P_{0}^{\mathbf{s}} \Phi_i^{\mathbf{s}} - \sum_{\mathbf{s}: s_i = 1} P_{0}^{\mathbf{s}} \Phi_i^{\mathbf{s}} \right)
\]

Replacing $\Phi_i^{\mathbf{s}}$, we get

\[
\mathcal{B} = \frac{1}{2N} \sum_{i=1}^N \sum_{j=1}^N a_{ij} \sum_{\mathbf{s}} P_{0}^{\mathbf{s}} (1 - s_i)s_j \left( \frac{1}{k_i} F[\beta(f^{\mathbf{s}}_i - f^{\mathbf{s}}_j)] - \frac{1}{k_j} F[-\beta(f^{\mathbf{s}}_i - f^{\mathbf{s}}_j)] \right) \tag{A8}
\]

Pairwise interactions with cumulative payoffs in the network
A usual way of computing the fitness is by summing the payoff a player gets from interacting with all their neighbors. In the traditional 2-player games, we associate state A with cooperation and B with defection. The payoff for mutual cooperation is $R$, for mutual defection is $P$, and when the players have different strategies, the cooperator receives $S$ and the defector $T$. We can write the fitness as

\[
f^{\mathbf{s}}_i = s_i \sum_{k=1}^N a_{ik}(R s_k + S(1 - s_k)) + (1 - s_i) \sum_{k=1}^N a_{ik}(T s_k + P(1 - s_k))
\]

And the fitness difference as

\begin{align*}
f^{\mathbf{s}}_i - f^{\mathbf{s}}_j 
&= \sum_{k=1}^N \left[ a_{ik}((P - S + R - T)s_i + (T - P)) - a_{jk}((P - S + R - T)s_j + (T - P)) \right] s_k \\
&\quad + k_i((S - P)s_i + P) - k_j((S - P)s_j + P)
\end{align*}

Notice that in Eq.(10), the fitness only needs to be calculated whenever $a_{ij} = 1$, $s_i = 1$, and $s_j = 0$; otherwise, the element is zero anyway, so we can write

\[
\Delta f_{ji}[n_{ij}^{\mathbf{s}}, n_i^{\mathbf{s}}, n_j^{\mathbf{s}}] \equiv f^{\mathbf{s}}_i - f^{\mathbf{s}}_j \big|_{s_i=0, s_j=1, a_{ij}=1}
= (T - P) n_{i}^{1\backslash j} - (R - S) n_{j}^{1\backslash i} + k_i P - k_j S + T - P
\]

where $n_{i}^{1\backslash j}$ is the number of neighbors of $i$ in state 1 that are not $j$. Notice that $i$ and $j$ share exactly

\[
\sigma_{ij} = \sum_{k=1}^N a_{ik} a_{jk}
\]

neighbors, of which $n_{ij}$ are labeled as those in state 1. Then, we can decompose the neighbors between those that are shared between $i$ and $j$ and those that are not, i.e.,

\[
n_{i}^{1\backslash j} = n_{ij} + n_i,
\]

where $n_i$ are the neighbors of $i$ that are not shared with $j$ (or $j$ itself) and are in state 1. Finally, we can write

\[
\Delta f_{ji}[n_{ij}^{\mathbf{s}}, n_i^{\mathbf{s}}, n_j^{\mathbf{s}}] = (T - P - R + S)n_{ij}^{\mathbf{s}} + (T - P)n_i^{\mathbf{s}} - (R - S)n_j^{\mathbf{s}} + k_i P - k_j S + T - P. \tag{A9}
\]

In the stationary distribution, $P_{0}^{\mathbf{s}}$, the states of individuals are independent, so $n_{ij}^{\mathbf{s}}, n_i^{\mathbf{s}}, n_j^{\mathbf{s}}$ are binomial samplings of subgroups of size $\sigma_{ij}, k_i - 1 - \sigma_{ij},$ and $k_j - 1 - \sigma_{ij}$, respectively, with probability of success $1/2$.

Plugging this into Eq.(10), we get

\[
\mathcal{B} = \frac{1}{2} \cdot \frac{1}{N} \sum_{i,j=1}^N a_{ij} \sum_{\mathbf{s}} P^{0}_{\mathbf{s}} (1 - s_i) s_j \left[ \frac{1}{k_i} F[\beta \Delta f_{ji}[n_{ij}^{\mathbf{s}}, n_i^{\mathbf{s}}, n_j^{\mathbf{s}}]] - \frac{1}{k_j} F[-\beta \Delta f_{ji}[n_{ij}^{\mathbf{s}}, n_i^{\mathbf{s}}, n_j^{\mathbf{s}}]] \right]
\]

\begin{align*}
\mathcal{B} 
&= \frac{1}{2} \cdot \frac{1}{N} \sum_{i,j=1}^N a_{ij} \left(\frac{1}{2}\right)^{k_i + k_j - \sigma_{ij}} 
\sum_{n_s=0}^{\sigma_{ij}} \sum_{n_1=0}^{k_i - 1 - \sigma_{ij}} \sum_{n_2=0}^{k_j - 1 - \sigma_{ij}} 
\binom{\sigma_{ij}}{n_s} \binom{k_i - 1 - \sigma_{ij}}{n_1} \binom{k_j - 1 - \sigma_{ij}}{n_2} \\
&\quad \times \left[ \frac{1}{k_i} F[\beta \Delta f_{ji}[n_s, n_1, n_2]] 
- \frac{1}{k_j} F[-\beta \Delta f_{ji}[n_s, n_1, n_2]] \right]\tag{A10}
\end{align*}

We can decompose this into two terms: one that contains the effects of the game under homogeneous networks, and another that contains the contribution of heterogeneity.

\[
\mathcal{B} = \mathcal{B}^{Ho}[\beta, a] + \mathcal{B}^{He}[\beta, a]
\]

with

% \[
% \mathcal{B}^{Ho}[\beta, a] = \frac{1}{2} \cdot \frac{1}{N} \sum_{i,j=1}^N a_{ij} \cdot \frac{1}{k_i} \times \ldots
% \]

\begin{align*}
\mathcal{B}^{Ho}[\beta, a] 
&= \frac{1}{2N} \sum_{i,j=1}^N a_{ij} \cdot \frac{1}{k_i}
\left(\frac{1}{2}\right)^{k_i + k_j - \sigma_{ij}} \sum_{n_s=0}^{\sigma_{ij}} 
\sum_{n_1=0}^{k_i - 1 - \sigma_{ij}} \sum_{n_2=0}^{k_j - 1 - \sigma_{ij}} \binom{\sigma_{ij}}{n_s} 
\binom{k_i - 1 - \sigma_{ij}}{n_1} \binom{k_j - 1 - \sigma_{ij}}{n_2} \\
&\quad \times \left[ F[\beta \Delta f_{ji}[n_s, n_1, n_2]] 
- F[-\beta \Delta f_{ji}[n_s, n_1, n_2]] \right] \tag{A11}
\end{align*}

\begin{align*}
\mathcal{B}^{He}[\beta, a] 
&= \frac{1}{2N} \sum_{i,j=1}^N a_{ij} \left(\frac{1}{k_i} - \frac{1}{k_j} \right)
\left(\frac{1}{2}\right)^{k_i + k_j - \sigma_{ij}} \\ & \times \sum_{n_s=0}^{\sigma_{ij}} 
\sum_{n_1=0}^{k_i - 1 - \sigma_{ij}} \sum_{n_2=0}^{k_j - 1 - \sigma_{ij}} 
\binom{\sigma_{ij}}{n_s} \binom{k_i - 1 - \sigma_{ij}}{n_1} \binom{k_j - 1 - \sigma_{ij}}{n_2} 
 F[-\beta \Delta f_{ji}[n_s, n_1, n_2]]\tag{A12}
\end{align*}

\section{The Fermi-update}

If we choose the traditional Fermi-update, with \( F(x) = (1 + e^{-x})^{-1} \), then, noticing that
\( F(x) - F(-x) = -\tanh\left(\frac{x}{2}\right) \) and \( F(-x) = 1 - F(x) \), we get

\begin{align*}
\mathcal{B}^{Ho}[\beta, a] 
&= -\frac{1}{2N} \sum_{i,j=1}^N a_{ij} \cdot \frac{1}{k_i}
\left(\frac{1}{2}\right)^{k_i + k_j - \sigma_{ij}} 
\sum_{n_s=0}^{\sigma_{ij}} \sum_{n_1=0}^{k_i - 1 - \sigma_{ij}} 
\sum_{n_2=0}^{k_j - 1 - \sigma_{ij}} 
\binom{\sigma_{ij}}{n_s} 
\binom{k_i - 1 - \sigma_{ij}}{n_1} 
\binom{k_j - 1 - \sigma_{ij}}{n_2} \\
&\quad \times \tanh\left( \frac{\beta}{2} \Delta f_{ji}[n_s, n_1, n_2] \right)
\end{align*}

\begin{align*}
\mathcal{B}^{He}[\beta, a] 
&= -\frac{1}{2N} \sum_{i,j=1}^N a_{ij} \left( \frac{1}{k_i} - \frac{1}{k_j} \right)
\left( \frac{1}{2} \right)^{k_i + k_j - \sigma_{ij}} \\
&\times \sum_{n_s=0}^{\sigma_{ij}} 
\sum_{n_1=0}^{k_i - 1 - \sigma_{ij}} 
\sum_{n_2=0}^{k_j - 1 - \sigma_{ij}} 
\binom{\sigma_{ij}}{n_s} 
\binom{k_i - 1 - \sigma_{ij}}{n_1} 
\binom{k_j - 1 - \sigma_{ij}}{n_2} 
 F\left[ \beta \Delta f_{ji}[n_s, n_1, n_2] \right]
\end{align*}

\subsection{Weak selection}

In the limit of weak selection (\( \beta \ll 1 \)), \( \tanh(\beta x) \approx \beta x \) and
\( F[x] \approx \frac{1}{2} - \frac{x}{4} \), ignoring terms in \( \beta^2 \). So, using Eq.(11), we get

\[
\mathcal{B} = \frac{\beta}{2^5} \left( P - T - R + S - (T + P)\bar{k} + (R + S)\overline{k_{nn}} +
\frac{1}{2}(T + P + R + S)(\bar{k} - \overline{k_{nn}}) \right)
\]

with \( \bar{k} \equiv \frac{1}{N} \sum_{i=1}^N k_i \) and
\( \overline{k_{nn}} \equiv \frac{1}{N} \sum_{i=1}^N \frac{1}{k_i} \sum_{j=1}^N a_{ij} k_j \). Or,

\[
\mathcal{B} = \frac{\beta}{2^5} (\kappa + 1) \left( \frac{\kappa - 1}{\kappa + 1} (R + S - T) - \frac{\kappa - 1}{\kappa + 1} P \right)
\]

with \( \kappa = \frac{\bar{k} + \overline{k_{nn}}}{2} \).

\subsection{Proof}

We start by showing that

\[
\mathcal{B}^{Ho}[\beta, a] \approx \frac{\beta}{2^5}(P - T - R + S - (T + P)\bar{k} + (R + S)\overline{k_{nn}})
\]

Using the definitions and the properties of the binomial distribution, we get

\begin{align*}
\mathcal{B}^{Ho}[\beta, a] 
&\approx -\frac{1}{2} \cdot \frac{1}{N} \cdot \beta \cdot \left(\frac{1}{2}\right)^{k_i + k_j - \sigma_{ij}} \cdot \frac{1}{k_i} \\
& \times\sum_{i=1}^N \sum_{j=1}^N a_{ij} 
\sum_{n_s=0}^{\sigma_{ij}} \sum_{n_1=0}^{k_i - 1 - \sigma_{ij}} 
\sum_{n_2=0}^{k_j - 1 - \sigma_{ij}} 
\binom{\sigma_{ij}}{n_s} 
\binom{k_i - 1 - \sigma_{ij}}{n_1} 
\binom{k_j - 1 - \sigma_{ij}}{n_2} \\
&\quad \times \Delta f_{ji}[n_s, n_1, n_2]\\
&= -\frac{1}{2} \cdot \frac{1}{N} \cdot \beta \cdot 
\sum_{i=1}^N \sum_{j=1}^N a_{ij} \left(\frac{1}{2}\right)^{k_i + k_j - \sigma_{ij}} \cdot \frac{1}{k_i}\\
& \times \sum_{n_s=0}^{\sigma_{ij}} \sum_{n_1=0}^{k_i - 1 - \sigma_{ij}} \sum_{n_2=0}^{k_j - 1 - \sigma_{ij}}
\binom{\sigma_{ij}}{n_s} \binom{k_i - 1 - \sigma_{ij}}{n_1} \binom{k_j - 1 - \sigma_{ij}}{n_2}\\
& \times\left[ (T - P - R + S)n_s + (T - P)n_1 - (R - S)n_2 + k_i P - k_j S + T - P \right]\\
&= -\frac{1}{2} \cdot \frac{1}{N} \cdot \beta \cdot \sum_{i=1}^N \sum_{j=1}^N a_{ij} \left(\frac{1}{2}\right)^{k_i + k_j - \sigma_{ij}} \cdot \frac{1}{k_i}\\
& \times\left(\left(\frac{1}{2}\right)^{3 - k_i - k_j + \sigma_{ij}} \left((T - P + R - S) + (T + P)k_i - (R + S)k_j\right)\right)\\
&= -\frac{1}{2} \cdot \beta \cdot \left(\frac{1}{2}\right)^3 \left((T - P + R - S) + (T + P)\bar{k} - (R + S)\overline{k_{nn}}\right)\\
&= \frac{\beta}{2^5}(P - T - R + S - (T + P)\bar{k} + (R + S)\overline{k_{nn}})\\
&\qquad \qquad \qquad \qquad \qquad \square
\end{align*}

Now we show that

\[
\mathcal{B}^{He}[\beta, a] \approx \frac{\beta}{4} \left(\frac{1}{2}\right)^4 (T + P + R + S)(\bar{k} - \overline{k_{nn}})
\]

Using definitions, properties of the binomial distribution, and noticing that
\(\sum a_{ij} b_{ij} = 0\) for any symmetric \(a\) and antisymmetric \(b\), we have

%\section*{Proof (continued)}
\begin{align*}
\mathcal{B}^{He}[\beta, a] 
&\approx -\frac{1}{2} \cdot \frac{1}{N} \sum_{i=1}^N \sum_{j=1}^N a_{ij}
\left( \frac{1}{2} \right)^{k_i + k_j - \sigma_{ij}} \left( \frac{1}{k_i} - \frac{1}{k_j} \right)\\
& \times\sum_{n_s=0}^{\sigma_{ij}} \sum_{n_1=0}^{k_i - 1 - \sigma_{ij}} 
\sum_{n_2=0}^{k_j - 1 - \sigma_{ij}} 
\binom{\sigma_{ij}}{n_s} 
\binom{k_i - 1 - \sigma_{ij}}{n_1} 
\binom{k_j - 1 - \sigma_{ij}}{n_2} 
\left( -\frac{\beta}{4} \Delta f_{ji}[n_s, n_1, n_2] \right) \\
&= -\frac{1}{2} \cdot \frac{1}{N} \cdot \frac{1}{2} \sum_{i,j=1}^N a_{ij}
\left( \frac{1}{2} \right)^{k_i + k_j - \sigma_{ij}} \left( \frac{1}{k_i} - \frac{1}{k_j} \right)\\
&\times\sum_{n_s=0}^{\sigma_{ij}} \sum_{n_1=0}^{k_i - 1 - \sigma_{ij}} 
\sum_{n_2=0}^{k_j - 1 - \sigma_{ij}} 
\binom{\sigma_{ij}}{n_s} 
\binom{k_i - 1 - \sigma_{ij}}{n_1} 
\binom{k_j - 1 - \sigma_{ij}}{n_2} 
\left( -\frac{\beta}{4} \Delta f_{ji}[n_s, n_1, n_2] \right) \\
&= \frac{\beta}{4} \cdot \frac{1}{N} \sum_{i,j=1}^N a_{ij}
\left( \frac{1}{2} \right)^{k_i + k_j - \sigma_{ij}} \left( \frac{1}{k_i} - \frac{1}{k_j} \right)\\
&\times\sum_{n_s=0}^{\sigma_{ij}} \sum_{n_1=0}^{k_i - 1 - \sigma_{ij}} 
\sum_{n_2=0}^{k_j - 1 - \sigma_{ij}} 
\binom{\sigma_{ij}}{n_s} 
\binom{k_i - 1 - \sigma_{ij}}{n_1} 
\binom{k_j - 1 - \sigma_{ij}}{n_2} 
\Delta f_{ji}[n_s, n_1, n_2] \\
&= \frac{\beta}{4} \cdot \frac{1}{N} \left( \frac{1}{2} \right)^3 
\sum_{i,j=1}^N a_{ij} \left( \frac{1}{k_i} - \frac{1}{k_j} \right) \left[ (T + P)k_i - (R + S)k_j \right] \\
&= \frac{\beta}{4} \left( \frac{1}{2} \right)^3 (T + P + R + S)
\sum_{i,j=1}^N a_{ij} \left( \frac{k_i}{k_i} - \frac{k_i}{k_j} \right) \\
&= \frac{\beta}{4} \left( \frac{1}{2} \right)^4 (T + P + R + S)(\bar{k} - \overline{k_{nn}})\\
&\qquad \qquad \qquad \qquad \qquad \square
\end{align*}

Putting the two together,

\[
\mathcal{B} = \frac{\beta}{2^5}(P - T - R + S - (T + P)\bar{k} + (R + S)\overline{k_{nn}}
+  \frac{1}{2} \cdot (T + P + R + S)(\bar{k} - \overline{k_{nn}}))
\]

\subsection{Strong selection}

For $\beta \gg 1$, 
\[
\tanh[\beta x] = \Theta[x] - \Theta[-x] = \Theta[x] - (1 - \Theta[x]) = 2\Theta[x] - 1,\quad
F[\beta x] = \Theta[-x],
\]
where $\Theta[x]$ is the Heaviside Theta function, that is 1 if $x > 0$ and 0 if $x < 0$, and modified to have $\Theta[0] = \frac{1}{2}$.

\[
\mathcal{B} = \mathcal{B}^{Ho}[\beta, a] + \mathcal{B}^{He}[\beta, a]
\]

with
\begin{align*}
\mathcal{B}^{Ho}[\beta, a] 
&\approx -\frac{1}{2} \cdot \frac{1}{N} \sum_{i=1}^N \sum_{j=1}^N a_{ij} 
\left( \frac{1}{2} \right)^{k_i + k_j - \sigma_{ij}} \cdot \frac{1}{k_i}\\
&\times\sum_{n_s=0}^{\sigma_{ij}} 
\sum_{n_1=0}^{k_i - 1 - \sigma_{ij}} 
\sum_{n_2=0}^{k_j - 1 - \sigma_{ij}} 
\binom{\sigma_{ij}}{n_s} 
\binom{k_i - 1 - \sigma_{ij}}{n_1} 
\binom{k_j - 1 - \sigma_{ij}}{n_2}
\left(2\Theta[\Delta f_{ji}[n_s, n_1, n_2]] - 1 \right) \\
&= -\frac{1}{2} \cdot \frac{1}{N} \sum_{i,j=1}^N a_{ij} 
\left( \frac{1}{2} \right)^{k_i + k_j - \sigma_{ij}} \cdot \frac{1}{k_i}\\
&\times\sum_{n_s=0}^{\sigma_{ij}} 
\sum_{n_1=0}^{k_i - 1 - \sigma_{ij}} 
\sum_{n_2=0}^{k_j - 1 - \sigma_{ij}} 
\binom{\sigma_{ij}}{n_s} 
\binom{k_i - 1 - \sigma_{ij}}{n_1} 
\binom{k_j - 1 - \sigma_{ij}}{n_2} 
\Theta[\Delta f_{ji}[n_s, n_1, n_2]] \\
&\quad + \frac{1}{2} \cdot \frac{1}{N} \sum_{i,j=1}^N a_{ij} 
\left( \frac{1}{2} \right)^{k_i + k_j - \sigma_{ij}} \cdot \frac{1}{k_i}
\sum_{n_s=0}^{\sigma_{ij}} 
\sum_{n_1=0}^{k_i - 1 - \sigma_{ij}} 
\sum_{n_2=0}^{k_j - 1 - \sigma_{ij}} 
\binom{\sigma_{ij}}{n_s} 
\binom{k_i - 1 - \sigma_{ij}}{n_1} 
\binom{k_j - 1 - \sigma_{ij}}{n_2} \\
&= -\frac{1}{2} \cdot \frac{1}{N} \sum_{i,j=1}^N a_{ij} 
\left( \frac{1}{2} \right)^{k_i + k_j - \sigma_{ij}} \cdot \frac{1}{k_i}
\sum_{n_s, n_1, n_2} 
\binom{\sigma_{ij}}{n_s} 
\binom{k_i - 1 - \sigma_{ij}}{n_1} 
\binom{k_j - 1 - \sigma_{ij}}{n_2} \\
&\quad \times \Theta\left[(T - P - R + S)n_s + (T - P)n_1 - (R - S)n_2 + k_i P - k_j S + T - P\right]
\end{align*}

\[
\mathcal{B}^{He}[\beta, a] \approx
\begin{aligned}[t]
& \frac{1}{2} \cdot \frac{1}{N} \sum_{i,j=1}^N a_{ij} \left( \frac{1}{k_i} - \frac{1}{k_j} \right)
\left( \frac{1}{2} \right)^{k_i + k_j - \sigma_{ij}} \\
&\times\sum_{n_s, n_1, n_2}
\binom{\sigma_{ij}}{n_s} 
\binom{k_i - 1 - \sigma_{ij}}{n_1} 
\binom{k_j - 1 - \sigma_{ij}}{n_2} 
\Theta[\Delta f_{ji}[n_s, n_1, n_2]]
\end{aligned}
\]

Where the term following appears in both $\mathcal{B}^{He}$ and $\mathcal{B}^{Ho}$ expressions:

\[
\mathcal{B} =
\begin{aligned}[t]
& -\frac{1}{2} \cdot \frac{1}{N} \sum_{i,j=1}^N a_{ij} 
\left( \frac{1}{2} \right)^{k_i + k_j - \sigma_{ij}} \cdot \frac{1}{k_i}
\sum_{n_s, n_1, n_2} 
\binom{\sigma_{ij}}{n_s} 
\binom{k_i - 1 - \sigma_{ij}}{n_1} 
\binom{k_j - 1 - \sigma_{ij}}{n_2} 
\Theta[\Delta f_{ji}[n_s, n_1, n_2]] \\
& + \frac{1}{2} \cdot \frac{1}{N} \sum_{i,j=1}^N a_{ij} 
\left( \frac{1}{k_i} - \frac{1}{k_j} \right)
\left( \frac{1}{2} \right)^{k_i + k_j - \sigma_{ij}} 
\sum_{n_s, n_1, n_2} 
\binom{\sigma_{ij}}{n_s} 
\binom{k_i - 1 - \sigma_{ij}}{n_1} 
\binom{k_j - 1 - \sigma_{ij}}{n_2} 
\Theta[\Delta f_{ji}[n_s, n_1, n_2]]
\end{aligned}
\]

\section{A particular game}

Let us consider a set of parameters common in the literature. For \( R = 1, P = 0, T = 1 - S \), we get

\[
\begin{aligned}[t]
\left(\frac{1}{2}\right)^{k_i + k_j - \sigma_{ij} - 2}
\sum_{n_s = 0}^{\sigma_{ij}} \sum_{n_1 = 0}^{k_i - 1 - \sigma_{ij}} \sum_{n_2 = 0}^{k_j - 1 - \sigma_{ij}} 
\binom{\sigma_{ij}}{n_s} 
\binom{k_i - 1 - \sigma_{ij}}{n_1} 
\binom{k_j - 1 - \sigma_{ij}}{n_2}\\
\times\Theta[(1 - S)n_1 - (1 - S)n_2 - k_j S + 1 - S]
\end{aligned}
\]
\[
= \left(\frac{1}{2}\right)^{k_i + k_j - 2\sigma_{ij} - 2}
\sum_{n_1 = 0}^{k_i - 1 - \sigma_{ij}} \sum_{n_2 = 0}^{k_j - 1 - \sigma_{ij}}
\binom{k_i - 1 - \sigma_{ij}}{n_1} \binom{k_j - 1 - \sigma_{ij}}{n_2}
\Theta[(1 - S)(n_1 - n_2) - k_j S + (1 - S)]
\]

The difference of two binomially distributed variables with parameter \( \frac{1}{2} \) is just a shifted binomial
distribution (with the shift of the mean of the subtracted variable), and sample given by the sum of the
two, and parameter still \( \frac{1}{2} \). I.e.,

\[
\left(\frac{1}{2}\right)^{k_i + k_j - 2\sigma_{ij} - 2}
\sum_{n_1 = 0}^{k_i - 1 - \sigma_{ij}} \sum_{n_2 = 0}^{k_j - 1 - \sigma_{ij}}
\binom{k_i - 1 - \sigma_{ij}}{n_1} \binom{k_j - 1 - \sigma_{ij}}{n_2}
g[n_1 - n_2]
\]

\[
= \left(\frac{1}{2}\right)^{k_i + k_j - 2\sigma_{ij} - 2}
\sum_{n_d = 0}^{k_i - 1 - \sigma_{ij} + k_j - 1 - \sigma_{ij}}
\binom{k_i + k_j - 2 - 2\sigma_{ij}}{n_d} g[n_d - k_j + 1 + \sigma_{ij}]
\]

So,

\[
\left(\frac{1}{2}\right)^{k_i + k_j - 2\sigma_{ij} - 2}
\sum_{n_1 = 0}^{k_i - 1 - \sigma_{ij}} \sum_{n_2 = 0}^{k_j - 1 - \sigma_{ij}}
\binom{k_i - 1 - \sigma_{ij}}{n_1} \binom{k_j - 1 - \sigma_{ij}}{n_2}
\Theta[(1 - S)(n_1 - n_2) - k_j S + (1 - S)]
\]

\[
= \left(\frac{1}{2}\right)^{k_i + k_j - 2\sigma_{ij} - 2}
\sum_{n_d = 0}^{k_i + k_j - 2 - 2\sigma_{ij}}
\binom{k_i + k_j - 2 - 2\sigma_{ij}}{n_d}
\Theta[(1 - S)n_d - k_j + (2 + \sigma_{ij})(1 - S)]
\]

If \( S \geq 1 \), this is necessarily zero. So, we can write, using the notation where

\[
\mathbf{1}_{\text{condition}} \equiv
\begin{cases}
1, & \text{condition is True} \\
0, & \text{condition is False}
\end{cases}
\]

\[
= \left( \frac{1}{2} \right)^{k_i + k_j - 2\sigma_{ij} - 2}
\sum_{n_d = \left\lceil \frac{k_j}{1 - S} - 2 - \sigma_{ij} \right\rceil}^{(k_i + k_j)(S < 1) - 2\sigma_{ij} - 2}
\binom{k_i + k_j - 2 - 2\sigma_{ij}}{n_d} \mathbf{1}_{S < 1}
\]

\[
= \mathbf{1}_{S < 1} \cdot \binom{k_i + k_j - 2 - 2\sigma_{ij}}{\left\lceil \frac{k_j}{1 - S} - 2 - \sigma_{ij} \right\rceil}
{}_2F_1\left[ k_i + k_j - 1 - 2\sigma_{ij};\ \left\lceil \frac{k_j}{1 - S} - 2 - \sigma_{ij} \right\rceil;\ \left\lceil \frac{k_j}{1 - S} - 1 - \sigma_{ij} \right\rceil;\ -1 \right]
\]

Thus,

\[
\mathcal{B}^{Ho}[\beta, a] \approx
\begin{aligned}[t]
& \frac{1}{2} \cdot \frac{1}{2^2} 
- \frac{1}{2} \cdot \mathbf{1}_{S < 1} \cdot \frac{1}{2} \cdot \frac{1}{N}
\sum_{i=1}^N \sum_{j=1}^N \frac{a_{ij}}{k_i} \binom{k_i + k_j - 2 - 2\sigma_{ij}}{\left\lceil \frac{k_j}{1 - S} - 2 - \sigma_{ij} \right\rceil} \\
&\quad \times {}_2F_1\left[ 
k_i + k_j - 1 - 2\sigma_{ij};\ 
\left\lceil \frac{k_j}{1 - S} - 2 - \sigma_{ij} \right\rceil;\ 
\left\lceil \frac{k_j}{1 - S} - 1 - \sigma_{ij} \right\rceil;\ 
-1 
\right]
\end{aligned}
\]

\[
\mathcal{B}^{He}[\beta, a] \approx
\begin{aligned}[t]
& \mathbf{1}_{S < 1} \cdot \frac{1}{2} \cdot \frac{1}{N} \cdot \frac{1}{2^2}
\sum_{i=1}^N \sum_{j=1}^N a_{ij} \left( \frac{1}{k_i} - \frac{1}{k_j} \right)
\binom{k_i + k_j - 2 - 2\sigma_{ij}}{\left\lceil \frac{k_j}{1 - S} - 2 - \sigma_{ij} \right\rceil} \\
&\quad \times {}_2F_1\left[
k_i + k_j - 1 - 2\sigma_{ij};\ 
\left\lceil \frac{k_j}{1 - S} - 2 - \sigma_{ij} \right\rceil;\ 
\left\lceil \frac{k_j}{1 - S} - 1 - \sigma_{ij} \right\rceil;\ 
-1
\right]
\end{aligned}
\]

Or, putting those together,

\[
\mathcal{B}^{Ho}[\beta, a] + \mathcal{B}^{He}[\beta, a] \approx
\begin{aligned}[t]
& \frac{1}{2^3}
- \mathbf{1}_{S < 1} \cdot \frac{1}{2^3} \cdot \frac{1}{N}
\sum_{i=1}^N \sum_{j=1}^N a_{ij} \left( \frac{3}{k_i} - \frac{1}{k_j} \right)
\binom{k_i + k_j - 2 - 2\sigma_{ij}}{\left\lceil \frac{k_j}{1 - S} - 2 - \sigma_{ij} \right\rceil} \\
&\quad \times {}_2F_1\left[
k_i + k_j - 1 - 2\sigma_{ij};\ 
\left\lceil \frac{k_j}{1 - S} - 2 - \sigma_{ij} \right\rceil;\ 
\left\lceil \frac{k_j}{1 - S} - 1 - \sigma_{ij} \right\rceil;\ 
-1
\right]
\end{aligned}
\]

\section{The weak dependence limit in human experiments}

Let us assume that a population evolves according to an update process described in experiments
performed with real people \cite{grujic2010social}.
In these experiments, a large group of individuals was identified as moody cooperators, and further,
they have identified the explicit dependence on the neighborhood and showed a base-line strategy
change rate for each strategy for randomized neighborhoods. Thus, we write the probabilities of
changing strategy (for cooperators and defectors) as a linear function of the number of cooperators:

\[
p^{\mathbf{s}}_{C,i} = \beta_C^0 + \gamma\left(\beta_C - \beta_C^0 + \alpha_C \sum_{j=1}^N a_{ij}s_j\right)
\]
\[
p^{\mathbf{s}}_{D,i} = \beta_D^0 + \gamma\left(\beta_D - \beta_D^0 + \alpha_D \sum_{j=1}^N a_{ij}s_j\right)
\]

Where:
\[
\beta_C = 1 - 0.413 = 0.587,\quad \alpha_C = -0.031,\quad \beta_C^0 = 1 - 0.490 = 0.510,
\]
\[
\beta_D = 0.254,\quad \alpha_D = 0.002,\quad \beta_D^0 = 0.202.
\]

The functions are such that when \( \gamma = 0 \), the probability of changing strategy matches the
outcome of the randomization of the neighborhoods in the experimental data (i.e., independent of the
number of neighbors). When \( \gamma = 1 \), it matches the behavior of the individuals in the first experiment.

Then, \( p^{\mathbf{s}}_{s_i,i} \equiv p_i^{(s_i)}[\gamma \Phi_i^{\mathbf{s}}] \), as

\[
p_i^{(s_i)}[\gamma \Phi_i^{\mathbf{s}}] =
s_i \left( \beta_C^0 + \gamma(\beta_C - \beta_C^0 + \alpha_C \sum_{j=1}^N a_{ij}s_j) \right)
+ (1 - s_i) \left( \beta_D^0 + \gamma(\beta_D - \beta_D^0 + \alpha_D \sum_{j=1}^N a_{ij}s_j) \right)
\]

\[
= s_i \beta_C^0 + (1 - s_i) \beta_D^0 + \gamma \left(
s_i (\beta_C - \beta_C^0 + \alpha_C \sum_{j=1}^N a_{ij}s_j) +
(1 - s_i)(\beta_D - \beta_D^0 + \alpha_D \sum_{j=1}^N a_{ij}s_j)
\right)
\]

\[
= s_i \beta_C^0 + (1 - s_i) \beta_D^0 + \gamma \Phi_i^{\mathbf{s}}
\]

And using equation (2), we get:

\[
\langle x \rangle = \frac{\beta_D^0}{\beta_C^0 + \beta_D^0}
+ \gamma \cdot \frac{1}{\beta_C^0 + \beta_D^0} \cdot \frac{1}{N}
\sum_{i=1}^N \left( \sum_{\mathbf{s}: s_i=0} P_{\mathbf{s}}^0 \Phi_i^{\mathbf{s}} - \sum_{\mathbf{s}: s_i=1} P_{\mathbf{s}}^0 \Phi_i^{\mathbf{s}} \right)
+ \mathcal{O}(\gamma^2)
\tag{A13}
\]

Now, we can compute the two sums as:

\[
\sum_{\mathbf{s}: s_i = 0} P_{\mathbf{s}}^0 \Phi_i^{\mathbf{s}} =
\frac{\beta_C^0}{\beta_C^0 + \beta_D^0}
\left( (\beta_D - \beta_D^0) + \alpha_D \cdot \frac{\beta_D^0}{\beta_C^0 + \beta_D^0} \cdot k_i \right)
\]

\[
\sum_{\mathbf{s}: s_i = 1} P_{\mathbf{s}}^0 \Phi_i^{\mathbf{s}} =
\frac{\beta_D^0}{\beta_C^0 + \beta_D^0}
\left( (\beta_C - \beta_C^0) + \alpha_C \cdot \frac{\beta_C^0}{\beta_C^0 + \beta_D^0} \cdot k_i \right)
\]

\[
\begin{aligned}
\langle x \rangle &= \frac{\beta_D^0}{\beta_C^0 + \beta_D^0}  \\
&\quad + \gamma \cdot \frac{1}{\beta_C^0 + \beta_D^0}
\left(
\frac{\beta_C^0(\beta_D - \beta_D^0)}{\beta_C^0 + \beta_D^0}
- \frac{\beta_D^0(\beta_C - \beta_C^0)}{\beta_C^0 + \beta_D^0}
+ (\alpha_D - \alpha_C) \cdot \frac{\beta_C^0}{\beta_C^0 + \beta_D^0} \cdot \frac{\beta_D^0}{\beta_C^0 + \beta_D^0}
\langle k \rangle \right) + \mathcal{O}(\gamma^2) 
\end{aligned}\tag{A14}
\]

Replacing values:

\[
\langle x \rangle = 0.284 + \gamma (0.0216 + 0.00589 \langle k \rangle) + \mathcal{O}(\gamma^2)
\]

Since the networks of all individuals are identical in the experiment (all individuals play in a square
lattice with their Moore’s neighbors), one cannot know whether, for heterogeneous networks, the
dependency is on the number of neighbors or on the fraction. An alternative response would be:

\[
p_{C,i}^{\mathbf{s}} = \beta_C^0 + \gamma \left(\beta_C - \beta_C^0 + 8\alpha_C \cdot
\frac{\sum_{j=1}^N a_{ij}s_j}{\sum_{j=1}^N a_{ij}}\right)
\]

\[
p_{D,i}^{\mathbf{s}} = \beta_D^0 + \gamma \left(\beta_D - \beta_D^0 + 8\alpha_D \cdot
\frac{\sum_{j=1}^N a_{ij}s_j}{\sum_{j=1}^N a_{ij}}\right)
\]

Where \( \beta_C = 1 - 0.413 = 0.587, \alpha_C = -0.031, \beta_C^0 = 1 - 0.490 = 0.510 \), and
\( \beta_D = 0.254, \alpha_D = 0.002, \beta_D^0 = 0.202 \).

In that case,

\[
\begin{aligned}
p_i^{(s_i)}[\gamma \Phi_i^{\mathbf{s}}] 
&= s_i \beta_C^0 + (1 - s_i) \beta_D^0 \\
&\quad + \gamma \left(
s_i \left( \beta_C - \beta_C^0 + 8\alpha_C \cdot \frac{\sum_{j=1}^N a_{ij}s_j}{\sum_{j=1}^N a_{ij}} \right)
+ (1 - s_i) \left( \beta_D - \beta_D^0 + 8\alpha_D \cdot \frac{\sum_{j=1}^N a_{ij}s_j}{\sum_{j=1}^N a_{ij}} \right)
\right)
\end{aligned}
\]

The average is still given by:

\[
\langle x \rangle = \frac{\beta_D^0}{\beta_C^0 + \beta_D^0}
+ \gamma \cdot \frac{1}{\beta_C^0 + \beta_D^0} \cdot \frac{1}{N}
\sum_{i=1}^N \left(
\sum_{\mathbf{s}: s_i = 0} P_{\mathbf{s}}^0 \Phi_i^{\mathbf{s}} -
\sum_{\mathbf{s}: s_i = 1} P_{\mathbf{s}}^0 \Phi_i^{\mathbf{s}}
\right) + \mathcal{O}(\gamma^2) \tag{A15}
\]

And we can compute the two sums as:

\[
\sum_{\mathbf{s}: s_i = 0} P_{\mathbf{s}}^0 \Phi_i^{\mathbf{s}} =
\frac{\beta_C^0}{\beta_C^0 + \beta_D^0} \left((\beta_D - \beta_D^0) + \alpha_D \cdot \frac{\beta_D^0}{\beta_C^0 + \beta_D^0}\right)
\]

\[
\sum_{\mathbf{s}: s_i = 1} P_{\mathbf{s}}^0 \Phi_i^{\mathbf{s}} =
\frac{\beta_D^0}{\beta_C^0 + \beta_D^0} \left((\beta_C - \beta_C^0) + \alpha_C \cdot \frac{\beta_C^0}{\beta_C^0 + \beta_D^0}\right)
\]

\[
\langle x \rangle = \frac{\beta_D^0}{\beta_C^0 + \beta_D^0}
+ \gamma \cdot \frac{1}{\beta_C^0 + \beta_D^0}
\left(
\frac{\beta_C^0(\beta_D - \beta_D^0)}{\beta_C^0 + \beta_D^0}
- \frac{\beta_D^0(\beta_C - \beta_C^0)}{\beta_C^0 + \beta_D^0}
+ 8(\alpha_D - \alpha_C) \cdot \frac{\beta_C^0 \beta_D^0}{(\beta_C^0 + \beta_D^0)^2}
\right) + \mathcal{O}(\gamma^2) \tag{A16}
\]
\end{widetext}
% \clearpage
\bibliography{apssamp}% Produces the bibliography via BibTeX.

\end{document}